\documentclass[12pt,a4paper]{article}

\usepackage[T1]{fontenc}
\usepackage{lmodern}
\usepackage{amsmath}
\usepackage{amsfonts}
\usepackage{amssymb}
\usepackage{mathrsfs}
\usepackage{physics}
\usepackage{graphicx}
\usepackage{caption}
\usepackage{subcaption}
\usepackage{jheppub}
\usepackage{enumitem}
\usepackage{framed,float}
\usepackage{booktabs}
\usepackage{array}
\usepackage{soul}
\usepackage{multirow}
\usepackage{comment}
\usepackage{braket}
\usepackage{graphicx}
\usepackage{todonotes}
\usepackage{youngtab}
\usepackage{tikz}
\usepackage{tabularx}
\usetikzlibrary{calc,arrows,decorations.markings}

\usepackage{color}

\preprint{}
\title{Quasinormal Modes of Extremal Reissner-Nordström Black Holes via Seiberg-Witten Quantization}

\author[a]{Yi-Rong Wang,}
\author[a]{Peng Yang,}
\author[a,b,c,1]{and Kilar Zhang\note{Corresponding Author.}}
\affiliation[a]{Department of Physics and Institute for Quantum Science and Technology, Shanghai University, 99 Shangda Road, Shanghai 200444, China}
\affiliation[b]{Shanghai Key Lab for Astrophysics, Shanghai Normal University, 100 Guilin Road, Shanghai 200234, China}
\affiliation[c]{Shanghai Key Laboratory of High Temperature Superconductors, Shanghai 200444, China}
\emailAdd{wyr2024@shu.edu.cn}
\emailAdd{yangpeng522@shu.edu.cn}
\emailAdd{kilar@shu.edu.cn}

\abstract{
We study the scalar perturbations of asymptotically flat extremal Reissner-Nordstr\"om black holes via the quantum Seiberg-Witten geometry of $\mathcal{N}=2$ SU(2) gauge theory with $N_f=2$ flavors. The radial master equation, governed by a double confluent Heun equation, is exactly mapped to the quantum Seiberg-Witten curve, providing an exact quantization condition derived from the non-perturbative Nekrasov-Shatashvili free energy. Analytically, this exact dictionary unveils precise gauge-theoretic interpretations for critical physical thresholds, demonstrating that the superradiance and mass decoupling limits naturally reduce the master equation to the Whittaker equation and the reduced doubly confluent Heun equation (the latter corresponds to the SW geometry of the $\mathcal{N}=2$ SU(2) gauge theory with $N_f=1$), respectively. At the strict extremal limit, the coalescence of horizons induces a topological singularity that complicates the spectral analysis. By accommodating this irregular singularity, our geometric framework resolves the singularity coalescence and enables the extraction of the discrete global quasinormal mode.  As our main contribution, we provide the first non-perturbative evaluation of the quasinormal modes spectrum for simultaneously charged and massive scalar fields directly at strict extremity. Furthermore, our analytical results reproduce numerical benchmarks for both neutral and charged massless probes, and naturally capture quasi-resonance behaviors.
}

\begin{document}
\maketitle
\flushbottom

\section{Introduction}
\label{sec:intro}

Since the seminal work of Seiberg and Witten in the 1990s \cite{Seiberg:1994rs,Seiberg:1994aj}, the study of four-dimensional $\mathcal{N}=2$ supersymmetric Yang-Mills (SYM) theories has served as a paradigmatic framework for understanding strong coupling dynamics in quantum field theory \cite{Argyres:1995jj, Minahan:1996fg}. By exploiting the constraints imposed by holomorphy and electromagnetic duality, Seiberg-Witten~(SW) theory successfully reduces the computation of the low-energy effective action to the geometric analysis of Riemann surfaces (SW curves) and their moduli spaces \cite{Klemm:1994qs,Donagi:1995cf}. This breakthrough not only unveiled the deep non-perturbative structure of gauge theories, further elucidated by Nekrasov's partition functions \cite{Nekrasov:2002qd}, but also established profound connections with classical integrable systems \cite{Gorsky:1995zq}, matrix models \cite{Mironov:2009uv}, and two-dimensional conformal field theories via the AGT correspondence \cite{Alday:2009aq,Wyllard:2009hg}.

In the realm of gravitational physics, black hole perturbation theory remains a cornerstone for probing spacetime geometry and testing the predictions of general relativity \cite{Regge:1957td,Zerilli:1970se}. The relaxation process of a perturbed black hole is characterized by its spectrum of complex frequencies, known as quasinormal mode~(QNM) \cite{Berti:2009kk,Konoplya:2011qq,Kokkotas:1999bd}. These modes act as the unique ``fingerprints'' of the black hole; their frequencies encode essential information about its mass, charge, and angular momentum, and are entirely independent of the initial perturbation.

With the advent of gravitational wave astronomy, the study of QNMs has transitioned from a theoretical abstraction to an observational reality.  The LIGO and Virgo collaborations have successfully detected the waveforms from binary black hole mergers, including the ``ringdown'' phase governed by QNM, the stage where the remnant black hole radiates away distortions via gravitational waves \cite{LIGOScientific:2016aoc,LIGOScientific:2016lio}. By analyzing the frequencies and damping times of these signals, physicists can perform ``black hole spectroscopy'' \cite{Dreyer:2003bv,Berti:2016lat} to rigorously test the No-Hair ``theorem'' \cite{Isi:2019aib}, and search for potential deviations from general relativity \cite{Yunes:2013dva,Berti:2018cxi}. For isolated black holes in asymptotically flat spacetimes, QNMs are fundamental observables that characterize their classical stability and late-time dynamical evolution \cite{Nollert:1999ji}.

Despite their physical importance, computing QNMs analytically is notoriously difficult. The perturbation equations typically take the form of highly coupled systems of differential equations with complex effective potentials. Traditional approaches often rely on semi-analytic techniques and numerical root-finding routines (e.g., Leaver's continued fraction method \cite{Leaver:1985ax}), or asymptotic approximations like the WKB method \cite{Iyer:1986np,Konoplya:2003ii}. While effective for numerical evaluation, these conventional methods 
do not directly reveal 
the underlying analytic structure of the eigenvalue problem, making it challenging to provide an exact, geometrically motivated quantization condition. Furthermore, these methods encounter degeneracies or precision loss when dealing with coalescing singularities, such as those inherently arising in the strict extremal black hole limit, hence making it difficult to smoothly track fundamental modes into the quasi-resonance regime. Very recently, exact WKB analysis has undergone significant refinements, successfully extracting the neutral massless spectrum at strict extremality with unprecedented precision~\cite{Hatsuda:2026ghx}. However, obtaining a unified, non-perturbative quantization condition for simultaneously charged and massive probes remains an open challenge.

Since 2020, a novel analytical framework bridging black hole perturbation theory and quantum SW geometry, named the SW/QNM correspondence, has been established \cite{Aminov:2020yma,Bianchi:2021xpr,Bianchi:2021mft,Hatsuda:2019eoj}. The mathematical foundation of this duality lies in the observation that the master equations governing black hole perturbations (e.g., the Teukolsky equation \cite{Teukolsky:1973ha}), upon appropriate transformation, can be mapped to the Heun class of differential equations \cite{Litvinov:2013sxa,Fiziev:2005ki}. 
Such analytic methods have also recently proven highly effective for computing exact holographic correlators \cite{He:2023wcs, He:2024xbi}.
These Heun equations are precisely the quantum versions of the Picard-Fuchs equations associated with SW curves. This correspondence provides a precise ``dictionary'' that maps black hole parameters to gauge theory quantities, transforming the gravitational eigenvalue problem into a solvable problem in quantum integrable systems using Nekrasov partition functions. In the gauge theory side, the hypermultiplets  with $N_f$ flavors  are associated with mass parameters $m = \{m_1, \dots, m_{N_f}\}$ that encode the physical properties of the black hole, such as its dimension and asymptotic geometry.  More recently, this framework is integrated into a broader classification of differential structures known as the extended Heun hierarchy, providing a systematic approach to the quantum geometry of Quiver SW theories~\cite{Ge:2025yqk, Yang:2026xrz}.

In this paper, we apply this analytic framework to the extremal Reissner-Nordstr\"om (RN) black hole in asymptotically flat spacetime \cite{,Senjaya:2024rse}. Extremal black holes are of particular interest in theoretical physics as they serve as zero-temperature ground states of charged black holes and represent the boundary of physically allowed states saturating the cosmic censorship bound. However, from the perspective of spectral analysis, the extremal limit is singular: the coalescence of the inner and outer horizons modifies the singularity structure of the underlying differential equation from a confluent Heun equation (CHE, typical for non-extremal cases) to a double confluent Heun equation~(DCHE). This structural change implies that the relevant gauge theory dual for the asymptotically flat extremal RN black hole is distinct from the $N_f=3$ or $N_f=4$ theories associated with standard Kerr, Kerr-de Sitter, or non-extremal RN geometries \cite{Gamayun:2013auu,Hatsuda:2020sbn}. Instead, it corresponds to the $\mathrm{SU(2)}$ SW theory with $N_f=2$ fundamental hypermultiplets.

We construct an exact dictionary for the scalar perturbation in the asymptotically flat extremal RN background, mapping it to the quantum SW curve of the $N_f=2$ theory. Based on this exact mapping, we uncover the precise gauge-theoretic interpretations of two critical physical thresholds: the superradiance limit ($\omega \to q$) and the mass decoupling limit ($\omega \to m_p$). We demonstrate that these physical limits trigger geometric reductions of the irregular singularities, forcing the master differential equation to degenerate into the Whittaker equation and the reduced doubly confluent Heun equation~(RDCHE), respectively. Remarkably, our SW/QNM dictionary flawlessly translates these geometric reductions into asymptotic limits of the gauge theory: the superradiance threshold corresponds to the strict weak-coupling limit where instanton corrections are completely suppressed, while the mass threshold exactly matches the canonical flavor-decoupling from the $N_f=2$ theory to the $N_f=1$ case.

Furthermore, we perform a non-perturbative computation of the QNMs frequencies for the neutral scalar ($q=0$) case, utilizing the exact quantization condition derived from the Nekrasov-Shatashvili (NS) free energy~\cite{Nekrasov:2009rc}. For the neutral massless case ($m_p=0$), we demonstrate that the instanton expansion (evaluated up to the 12-instanton order) yields analytical predictions that are in agreement with high-precision numerical benchmarks~\cite{Onozawa:1995vu}. For neutral massive fields ($m_p \neq 0$), our framework allows  to smoothly track the fundamental mode into the quasi-resonance regime ($\text{Im}(\omega) \to 0$), resolving the singularity coalescence that plagues traditional methods at the strict extremal limit. More importantly, for charged and massive fields ($q \neq 0, m_p \neq 0$), previous studies have utilized the continued fraction method to evaluate the QNMs spectrum in the near-extremal regime~\cite{Richartz:2015saa}~($q \neq 0, m_p = 0$), and 3rd-order WKB approximations provided estimations solely for the imaginary parts (damping rates) at certain specific parameter values~\cite{Konoplya:2002wt}~($q \neq 0, m_p \neq 0$). However, as the black hole gets closer to strict extremality, the numerical data tends to exhibit a certain degree of instability due to the coalescence of horizons. By intrinsically mapping the exact radial equation to the DCHE, our $N_f=2$ SW geometric framework naturally circumvents this mathematical difficulty. Consequently, we provide the first 
predictions for the discrete global QNMs of charged massive fields at the strict extremal limit. Finally, we note that our current perturbative framework exhibits numerical instabilities for certain parameter combinations. This constitutes an intriguing open challenge that warrants further investigation in future studies.

The paper is organized as follows. In Section~\ref{sec:sw_geo}, we review the quantum geometry of the SW theory and the associated quantization conditions. Section~\ref{sec:ern_pert} systematically constructs the dictionary mapping the extremal RN black hole scalar perturbations to the quantum SW curve, providing precise gauge-theoretic interpretations for the superradiance and mass decoupling limits. Section~\ref{sec:numerical} presents the numerical evaluation of the QNMs frequencies, covering the neutral massless, neutral massive, charged massless, and charged massive cases. We first validate our framework by demonstrating that our results for both neutral and charged massless scalars are in  agreement with existing numerical benchmarks. We also track the quasi-resonance behavior for the neutral massive case. Subsequently, as our main novel contribution, we provide the first 
predictions for the discrete QNMs spectrum of simultaneously charged and massive fields in the strictly extremal RN background. Finally, Section~\ref{sec:conclusion} provides a summary and outlines future research directions. Additionally, Appendix~\ref{app:ns} details the definition and expansion of the NS free energy, while Appendix~\ref{app:large} furnishes the comprehensive numerical data underlying Table~\ref{tab:charged_grid} (in Section~\ref{sec:charged_massive}), further substantiating our 
predictions for the charged massive perturbations.

\section{Quantum Seiberg-Witten Geometry with $N_f=2$}
\label{sec:sw_geo}

In this section, we briefly review the essential dictionary relating SW theory to spectral theory, following \cite{Aminov:2020yma}. We outline the geometric framework emerging from the NS limit, focusing specifically on the $N_f=2$ theory relevant for the extremal RN black hole.
Central to this framework is the concept of \textit{quantum periods}, which generalize the classical SW geometry to the quantum regime, thereby allowing for the exact computation of discrete spectral frequencies.

Consider a quantum operator $H(\hat{x},\hat{p})$ satisfying the canonical commutation relation $[\hat{x},\hat{p}]=i\hbar$. Associated with this operator is a classical curve defined by $H(x,p)=E$ and a differential one-form $\lambda=p\, \mathrm{d}x$. The classical periods are defined by integrating $\lambda$ along the non-trivial cycles of the curve, analogous to the SW periods in $\mathcal{N}=2$ gauge theory. To transit to the quantum regime, these classical periods are promoted to quantum periods via the WKB expansion:
\begin{equation}
    \Pi^{\mathrm{WKB}}(E,\hbar)=\sum_{n\geq 0}\hbar^{n}\oint_{\gamma} Q_{n}(x,E)\, \mathrm{d}x.
    \label{eq:wkb_period}
\end{equation}
Here, the functions $Q_{n}(x,E)$ represent the expansion coefficients of the quantum differential $\lambda(x,E,\hbar) = \sum_{n\geq 0}\hbar^{n}Q_{n}(x,E)\mathrm{d}x$. This quantum differential is formally constructed such that the exponential of its integral yields the wavefunction satisfying the quantum spectral equation, $(H(\hat{x},\hat{p})-E)\exp[\frac{i}{\hbar}\int^{x}\lambda(y,E,\hbar)]=0$ . Since the standard WKB series is typically asymptotic and divergent, a non-perturbative definition is required. The exact quantum periods are non-perturbatively determined by the NS free energy of the corresponding supersymmetric gauge theory \cite{Nekrasov:2009rc, Mironov:2009uv}. For the $\mathrm{SU(2)}$ theory, the quantum periods $\Pi_A^{(N_f)}$ and $\Pi_B^{(N_f)}$ associated with the A-cycle and B-cycle are given by:
\begin{equation}
\begin{aligned}
    \Pi_{A}^{(N_f)}(E,\mathbf{m},\Lambda_{N_{f}},\hbar) &= a(E,\mathbf{m},\Lambda_{N_{f}},\hbar), \\
    \Pi_{B}^{(N_f)}(E,\mathbf{m},\Lambda_{N_{f}},\hbar) &= \frac{\partial \mathcal{F}^{(N_{f})}(a,\mathbf{m},\Lambda_{N_{f}},\hbar)}{\partial a}\Bigg|_{a=a(E,\mathbf{m},\Lambda_{N_{f}},\hbar)}.
\end{aligned}
\label{eq:quantum_periods}
\end{equation}
Here, $a$ is the scalar vacuum expectation value (Coulomb branch parameter), $\mathbf{m}$ denotes the hypermultiplet masses, and $\Lambda_{N_f}$ is the dynamical scale of the theory. The function $\mathcal{F}^{(N_f)}$ is the NS free energy, which can be computed exactly using instanton counting techniques~\cite{Nekrasov:2009rc}. For the explicit combinatorial definitions and the instanton expansion for the $N_f=2$ case, we refer the reader to Appendix \ref{app:ns}.

Consequently, the exact quantization condition for the spectral problem is formulated as a Bohr-Sommerfeld-like condition involving the quantum B-period:
\begin{equation}
    \Pi_{B}^{(N_f)}(E,\mathbf{m},\Lambda_{N_{f}},\hbar)=\mathcal{N}_{B}\hbar\left(n+\frac{1}{2}\right), \quad n=0,1,2,\dots,
    \label{eq:quantization_cond}
\end{equation}
where $\mathcal{N}_B$ is a numerical constant depending on the specific normalization of the cycles ($2\pi$ in our convention).

This formalism provides a unified framework for various spectral problems, where the quantum SW curves for $\mathrm{SU(2)}$ gauge theories with $N_f$ fundamental hypermultiplets naturally map to specific families of Heun equations~(HE). While the initial finding  \cite{Aminov:2020yma} already analyzed the $N_f=2$ (DCHE) case in the extremal limit of the Kerr black hole, subsequent literature has predominantly focused on the extensive exploration of the $N_f=3$ (CHE) and $N_f=4$ (General Heun) scenarios (see e.g.,~\cite{Bonelli:2021uvf,Ge:2024jdx,Lei:2023mqx}). 
In addition to the Kerr black hole, the quantum Seiberg-Witten/gravity correspondence for the SU(2) gauge theory with $N_f=2$ fundamental hypermultiplets has also been successfully applied to study the tidal resonances of D1D5 and D1D5P fuzzballs \cite{DiRusso:2024hmd}. 
To the best of our knowledge, the exact correspondence between the $N_f=2$ gauge theory and the extremal RN black hole remains unexplored. In this paper, we bridge this gap by establishing this exact mapping and providing the first analytic computation of quasinormal modes for the extremal RN geometry using this framework. The quantum curve for the $N_f=2$ theory takes the form of the DCHE. It can be written as a Schrödinger-like equation:
\begin{equation}
    \hbar^{2}\Psi''(z)+\widehat{Q}_{2}(z)\Psi(z)=0,
    \label{eq:DCHE}
\end{equation}
with the potential given by:
\begin{equation}
    \widehat{Q}_{2}(z)=-\frac{\Lambda_{2}^{2}}{16}-\frac{m_{1}\Lambda_{2}}{2z}+\frac{4E + \hbar^{2}}{4z^{2}}-\frac{m_{2}\Lambda_{2}}{2z^{3}}-\frac{\Lambda_{2}^{2}}{16z^{4}}.
    \label{eq:Q2_potential}
\end{equation}
    This geometric/gauge-theoretic perspective not only resums the divergent WKB series but also provides the necessary tools to analyze the QNMs of black holes that reduce to this specific differential equation. The potential \eqref{eq:Q2_potential} possesses two irregular singular points at $z=0$ and $z=\infty$, mirroring the singularity structure of the extremal RN radial equation which we will discuss in the next section.

\section{Scalar Perturbations of Extremal RN Black Holes}
\label{sec:ern_pert}

In this section, we explicitly construct the dictionary relating the scalar perturbation of an extremal RN black hole to the quantum SW curve of the $\mathrm{SU(2)}$ gauge theory with $N_f=2$ fundamental hypermultiplets. We first reduce the radial wave equation to the standard DCHE and subsequently match the physical parameters to the gauge theory parameters.

\subsection{Scalar Perturbations in the Extremal Background}

The background metric of an extremal RN black hole, where the mass $M$ equals the electric charge $Q$, in standard spherical coordinates is given by
\begin{equation}
    \mathrm{d}s^{2} = -f(r)\mathrm{d}t^{2} + \frac{\mathrm{d}r^{2}}{f(r)} + r^{2}(\mathrm{d}\theta^{2} + \sin^2\theta \mathrm{d}\phi^{2}),
\end{equation}
where the metric function possesses a double root at the degenerate horizon $r=Q$,
\begin{equation}
    f(r) = \left(1-\frac{Q}{r}\right)^{2} = \frac{(r-Q)^{2}}{r^{2}}.
\end{equation}
The background electromagnetic gauge potential $A_{\mu}$ satisfies $A_{\mu}\mathrm{d}x^{\mu} = -\frac{Q}{r}\mathrm{d}t$. We consider a massive charged scalar field $\Phi$ with mass $m_{p}$ and charge $q$, governed by the covariant Klein-Gordon equation 
\begin{equation}
    (\nabla_{\mu} - iqA_{\mu})(\nabla^{\mu} - iqA^{\mu})\Phi - m_{p}^{2}\Phi = 0.
\end{equation}
Adopting the standard separation of variables $\Phi = e^{-i\omega t}R(r)Y_{lm}(\theta,\phi)$, where $Y_{lm}(\theta,\phi)$ represents the spherical harmonic function, so the radial wave equation becomes:
\begin{equation}
    \frac{\mathrm{d}}{\mathrm{d}r}\left[ (r-Q)^{2}\frac{\mathrm{d}}{\mathrm{d}r} \right]R + \left[ \frac{r^{2}(\omega r - qQ)^{2}}{(r-Q)^{2}} - l(l+1) - m_{p}^{2}r^{2} \right] R = 0.
\end{equation}

\subsection{Reduction to the Double Confluent Heun Equation}

To map this radial equation to the normal form of the HE, we first shift the radial coordinate to the horizon by defining $x = r - Q$. Subsequently, we redefine the radial function to eliminate the first derivative term by setting $R(x) = \frac{1}{x}Y(x)$. This yields a Schr\"odinger-like equation:
\begin{equation}
    \frac{\mathrm{d}^{2}Y}{\mathrm{d}x^{2}} + Q(x)Y = 0,
    \label{eq: the radial equation}
\end{equation}
where the effective potential $Q(x)$ is given by 
\begin{equation}
    Q(x) = \frac{(x+Q)^{2}[\omega(x+Q) - qQ]^{2}}{x^{4}} - \frac{m_{p}^{2}(x+Q)^{2}}{x^{2}} - \frac{l(l+1)}{x^{2}}.
    \label{eq: radial  effective potential}
\end{equation}
Expanding $Q(x)$ in inverse powers of $x$, we introduce a dimensionless coordinate $z = kx$ through the scaling factor 
\begin{equation}
    k = \frac{1}{Q} \left( \frac{\omega^{2} - m_{p}^{2}}{(\omega-q)^{2}} \right)^{\frac{1}{4}}.
\end{equation}
Assuming the physically relevant regime where the frequencies are non-trivial, this scaling remains well-defined.
The radial equation then assumes the standard form of the DCHE:
\begin{equation}
    \frac{\mathrm{d}^{2}Y}{\mathrm{d}z^{2}} + \left( A_{0} + \frac{A_{1}}{z} + \frac{A_{2}}{z^{2}} + \frac{A_{3}}{z^{3}} + \frac{A_{4}}{z^{4}} \right) Y(z) = 0,
\end{equation}
where the coefficients $A_i$ are exact algebraic functions of the physical black hole parameters:
\begin{equation}
\begin{aligned} 
    A_{0} &= Q^{2}(\omega-q)(\omega^{2}-m_{p}^{2})^{\frac{1}{2}}, \\ 
    A_{1} &= 2Q^{2}(2\omega^{2}-\omega q-m_{p}^{2})\left( \frac{(\omega-q)^{2}}{\omega^{2}-m_{p}^{2}} \right)^{\frac{1}{4}}, \\ 
    A_{2} &= Q^{2}(6\omega^{2}-6\omega q+q^{2}-m_{p}^{2}) - l(l+1), \\ 
    A_{3} &= 2Q^{2}(2\omega-q)(\omega-q)^{\frac{1}{2}}\left( \omega^{2}-m_{p}^{2} \right)^{\frac{1}{4}}, \\ 
    A_{4} &= Q^{2}(\omega-q)(\omega^{2}-m_{p}^{2})^{\frac{1}{2}}. 
\end{aligned}
\end{equation}

By matching the algebraic coefficients $A_i$ of the radial equation with the potential $\widehat{Q}_2(z)$ \eqref{eq:Q2_potential} of the quantum SW curve for the $N_f=2$ theory \eqref{eq:DCHE}, and setting the quantum deformation parameter $\hbar=1$ without loss of generality, we obtain a system of equations for the gauge theory parameters. In this work, we present the following branch as a consistent choice for our parameter dictionary:
\begin{equation}
\begin{aligned}
    \Lambda_{2} &= -4iQ \sqrt{\omega-q} (\omega^{2}-m_{p}^{2})^{\frac{1}{4}}, \\
    m_{1} &= -\frac{iQ(2\omega^{2}-\omega q-m_{p}^{2})}{\sqrt{\omega^{2}-m_{p}^{2}}}, \\
    m_{2} &= -iQ(2\omega-q), \\
    E &= Q^{2}(6\omega^{2}-6\omega q+q^{2}-m_{p}^{2}) - l(l+1) - \frac{1}{4}.
\end{aligned}
\label{eq:negative_branch}
\end{equation}

\subsection{Two Special Limits and Gauge Theory Interpretations}
\label{sec:special_limits}

The exact SW/QNM dictionary constructed in~\eqref{eq:negative_branch} not only facilitates numerical computations but also offers profound gauge-theoretic interpretations for specific physical limits of the black hole. By rigorously tracking the mathematical degeneration of the radial equation, we unveil the elegant correspondence between the geometric singularity reduction and the suppression of gauge instantons.

\subsubsection{The Superradiance Threshold ($\omega \to q$)}

For an extremal RN black hole, the electrostatic potential at the degenerate horizon $r_+ = Q$ is $\Phi_H = Q/r_+ = 1$. The threshold for superradiant scattering occurs when the frequency of the incoming scalar wave strictly matches the energy it can extract from the black hole, yielding the critical frequency $\omega = q \Phi_H = q$. 

To elucidate the mathematical mechanism underpinning this physical limit, let us revisit the radial effective potential $Q(x)$ defined in~\eqref{eq: radial  effective potential}. The original differential equation~\eqref{eq: the radial equation}, is a DCHE, characterized by a highly divergent irregular singularity of Poincaré rank 1 at the horizon ($x \to 0$) due to the $\mathcal{O}(x^{-4})$ and $\mathcal{O}(x^{-3})$ terms. However, in the strict superradiance threshold $\omega \to q$, a remarkable algebraic cancellation occurs inside the square bracket: $[\omega(x + Q) - qQ] \to qx$. Consequently, the leading divergence in the potential simplifies drastically:
\begin{equation}
    Q(x) \xrightarrow{\omega \to q} \frac{(x + Q)^2(qx)^2}{x^4} - \frac{m_p^2(x + Q)^2}{x^2} - \frac{l(l+1)}{x^2}.
\end{equation}
Factoring out the surviving terms, the effective potential strictly collapses to:
\begin{equation}
    Q(x) \to (q^2 - m_p^2) + \frac{2Q(q^2 - m_p^2)}{x} + \frac{Q^2(q^2 - m_p^2) - l(l+1)}{x^2}. \label{eq:Qx_collapsed}
\end{equation}
Mathematically, the $1/x^4$ and $1/x^3$ severe divergences have completely vanished. The irregular singularity at the horizon degenerates into a regular singularity. 

This reduction implies that the DCHE flows to the simpler Confluent Hypergeometric Equation. To manifest this, we define the asymptotic momentum scale $\alpha = \sqrt{m_p^2 - q^2}$ and perform a conformal coordinate scaling $z = 2\alpha x$. The radial equation then exactly maps to the standard Whittaker equation:
\begin{equation}
    \frac{d^2Y}{dz^2} + \left( -\frac{1}{4} + \frac{\kappa}{z} + \frac{1/4 - \mu^2}{z^2} \right)Y(z) = 0,
    \label{eq:Whittaker equation}
\end{equation}
where the Whittaker parameters $\kappa$ and $\mu$ are identified as:
\begin{equation}
    \kappa = -Q\alpha = -Q\sqrt{m_p^2 - q^2}, \quad \mu = \sqrt{\left(l+\frac{1}{2}\right)^2 + Q^2(m_p^2 - q^2)}.
\end{equation}
The physically acceptable wavefunction that respects the boundary condition at spatial infinity ($z \to \infty$) is uniquely given by the Whittaker $W$ function, $Y(z) \propto W_{\kappa, \mu}(z)$. To impose the physical boundary condition at the regular horizon ($z \to 0$), we utilize the connection formula relating $W_{\kappa, \mu}(z)$ to the fundamental solutions $M_{\kappa, \pm\mu}(z)$:
\begin{equation}
    W_{\kappa, \mu}(z) = \frac{\Gamma(-2\mu)}{\Gamma\left(\frac{1}{2} - \mu - \kappa\right)} M_{\kappa, \mu}(z) + \frac{\Gamma(2\mu)}{\Gamma\left(\frac{1}{2} + \mu - \kappa\right)} M_{\kappa, -\mu}(z).
\end{equation}
Selecting the physically regular branch at the horizon requires the coefficient of the divergent singular mode to vanish. This condition is naturally satisfied when the denominator hits a pole of the Gamma function, which occurs at non-positive integers. Thus, the exact analytical quantization condition is simply:
\begin{equation}
    \frac{1}{2} \pm \mu - \kappa = -n, \quad n = 0, 1, 2, \dots
\end{equation}
Substituting $\kappa$ and $\mu$ back yields a closed-form algebraic spectrum:
\begin{equation}
    \frac{1}{2} \pm \sqrt{\left(l+\frac{1}{2}\right)^2 + Q^2(m_p^2 - q^2)} + Q\sqrt{m_p^2 - q^2} = -n. \label{eq:whittaker_quantization}
\end{equation}

From the gauge theory perspective, this exact Gamma-function solvability finds a precise interpretation. Evaluating our exact dictionary~\eqref{eq:negative_branch} in the $\omega \to q$ limit, the dynamical scale vanishes, $\Lambda_2 \propto \sqrt{\omega - q} \to 0$, while the mass parameters remain finite constants ($m_1 \to -iQ\sqrt{q^2 - m_p^2}$ and $m_2 \to -iQq$). In $\mathcal{N}=2$ supersymmetric gauge theories, the instanton counting parameter is proportional to $\Lambda_2$. The condition $\Lambda_2 \to 0$ strictly confines the theory to the \textit{weak-coupling limit}. Consequently, all instanton corrections are completely suppressed ($\mathcal{F}_{\text{inst}} \to 0$). The quantization condition is entirely dictated by the perturbative prepotential (classical tree-level plus 1-loop contributions). This exactly explains why the superradiance critical condition $\omega \to q$~\eqref{eq:Whittaker equation} can be analytically resolved purely via Gamma functions, completely bypassing the infinite Nekrasov instanton resummation.

\subsubsection{The Mass Decoupling Limit ($\omega \to m_p$)}

Another fascinating physical regime occurs when the frequency of the scalar field approaches its rest mass, $\omega \to m_p$. Physically, this threshold marks the onset of quasi-bound states (or the quasi-resonance regime), where the asymptotic momentum of the scalar wave at spatial infinity vanishes, $k_\infty = \sqrt{\omega^2 - m_p^2} \to 0$. 

According to our dictionary~\eqref{eq:negative_branch}, in the limit $\omega \to m_p$, the gauge theory dynamical scale vanishes as $\Lambda_2 \to 0$, while the hypermultiplet mass parameter diverges as $m_1 \to \infty$. Remarkably, in this situation, the following equation still holds true:
\begin{equation}
    m_1 \Lambda_2^2 = 16 i Q^3 (\omega - q) (2\omega^2 - \omega q - m_p^2)= 16 i Q^3 m_p(m_p-q)^2.
\end{equation}
In the strict limit $\omega \to m_p$, this product converges to a well-defined finite constant $\Lambda_1^3 \equiv 16 i Q^3 m_p (m_p - q)^2$. In the context of Seiberg-Witten theory, this is the canonical decoupling limit where the $N_f=2$ gauge theory flows to the pure $N_f=1$ theory, with $\Lambda_1$ serving as its dynamical scale and $m_2$ remaining as the mass of the single surviving flavor.

To rigorously demonstrate this geometric flow, we perform a conformal coordinate scaling $z = \frac{\Lambda_2}{2} t$ on the original $N_f=2$ quantum SW curve (the DCHE). Using the transformation property of the quadratic differential, the new effective potential becomes $\tilde{Q}_1(t) = \widehat{Q}_2(z(t)) (\frac{dz}{dt})^2$. Then taking the $\Lambda_2 \to 0$ limit while holding $\Lambda_1^3 = m_1 \Lambda_2^2$ fixed, the potential collapses exactly to:
\begin{equation}
    \tilde{Q}_1(t) = -\frac{\Lambda_1^3}{4t} + \frac{4E+\hbar^2}{4t^2} - \frac{m_2}{t^3} - \frac{1}{4t^4}. \label{eq:rdche_potential}
\end{equation}
Geometrically, the vanishing asymptotic momentum reduces the Poincaré rank of the irregular singularity at spatial infinity. The underlying equation successfully degenerates from the DCHE to the RDCHE.

Astoundingly, this resulting RDCHE is mathematically isomorphic to the established $\mathcal{N}=2$ $\mathrm{SU(2)}$ quantum SW curve for the $N_f=1$ SQCD. To prove this, we recall that the quantum curve for the $N_f=1$ theory is canonically given by a Schrödinger equation (see equation (2.10) and equation (3.1) in \cite{Ito:2017iba}):
\begin{equation}
\begin{aligned}
    &\left( - \hbar^2 \partial_v^2 + Q(v) \right) \psi(v) = 0, \\
    &Q(v) = -\frac{1}{16} \Lambda_1^3 e^{2i v} - \frac{1}{2} \Lambda_1^{3/2} m e^{i v} - u - \frac{1}{2} \Lambda_1^{3/2} e^{-i v},
\end{aligned}
\end{equation}
where $u = -E$ is the Coulomb modulus and $m$ is the hypermultiplet mass. Setting $y = e^{iv}$ and redefining the wavefunction as $\Psi(y) = y^{-1/2}\psi(y)$ to eliminate the first-order derivative, the potential transforms into:
\begin{equation}
\begin{aligned}
    &\left(  \hbar^2 \partial_y^2 + Q(y) \right) \Psi(y) = 0, \\
    &\hat{Q}(y) = - \frac{\Lambda_1^3}{16} - \frac{m\Lambda_1^{3/2}}{2y} + \frac{\hbar^2 + 4E}{4y^2} - \frac{\Lambda_1^{3/2}}{2y^3}.
\end{aligned}
\label{eq:nf=1 sw curve}
\end{equation}
Thus, the rescaled wavefunction $\Psi(y)$ identically satisfies a Schrödinger-like equation with the effective potential $\hat{Q}(y)$.

As for our extremal RN black hole, in the mass limit condition, the potential is shown in~\eqref{eq:rdche_potential}. To align the singularity structures with~\eqref{eq:nf=1 sw curve}, we introduce a transformation $t=\frac{2}{\Lambda_1^{3/2} y}$and redefining the wavefunction as $\Psi(t) = t^{-1}Y(t)$ to eliminate the first-order derivative. The potential then transforms as $\tilde{Q}(t) = \hat{Q}(y(t)) \left(\frac{dy}{dt}\right)^2$. Since $\left(\frac{dy}{dt}\right)^2 = \frac{4}{\Lambda_1^3 t^4}$, we find:
\begin{equation}
    \tilde{Q}(t) = \left[ - \frac{1}{16}\Lambda_1^3 - \frac{1}{4}\Lambda_1^3 m t - \frac{1}{16}\Lambda_1^6 t^3 + \frac{\Lambda_1^3 t^2(\hbar^2 + 4E)}{16}\right] \frac{4}{\Lambda_1^3 t^4 }.
\end{equation}
Expanding this product precisely yields:
\begin{equation}
    \tilde{Q}(t) = -\frac{\Lambda_1^3}{4t} + \frac{\hbar^2 + 4E}{4t^2} - \frac{m}{t^3} - \frac{1}{4t^4}.
\end{equation}
Comparing this gauge-theoretic outcome with~\eqref{eq:rdche_potential}, the exact equivalence is manifest under the identification $m \leftrightarrow m_2$. This rigorous mathematical mapping encodes the mass decoupling limit (with $\Lambda_1^3 = m_1 \Lambda_2^2$) of the extremal RN black hole  into the canonical decoupling flow from the $N_f=2$ to the $N_f=1$ supersymmetric gauge theory. Consequently, the exact SW/QNM dictionary for the $N_f=1$ theory at the mass decoupling limit can be systematically deduced from this exact correspondence.

\subsubsection{Summary of the Two Limits.}
In summary, the exploration of these two special limits elegantly showcases the profound triality among black hole physics, the extended Heun hierarchy, and quantum gauge theories. From the gravitational perspective, $\omega \to q$ and $\omega \to m_p$ represent the critical physical thresholds defining the boundaries of black hole stability: the onset of superradiant instability and the cutoff of quasi-bound state dissipation, respectively. Geometrically, both limits trigger a topological reduction of the severe irregular singularity at the horizon or spatial infinity, forcing the DCHE to degenerate into simpler differential structures (the Whittaker equation and RDCHE). Most remarkably, the SW/QNM dictionary can translate these geometric reductions into canonical asymptotic limits of the $\mathcal{N}=2$ gauge theory: the superradiance threshold corresponds to the strict weak-coupling limit (where instantons are completely suppressed), while the mass threshold matches the canonical flavor-decoupling limit ($N_f=2 \to N_f=1$). This exact matching demonstrates that the SW/QNM framework is not merely a numerical tool, but a fundamental dictionary encoding black hole criticality into gauge-theoretic degenerations.

\section{Evaluation and Discussion}
\label{sec:numerical}

Having established the exact quantization condition via the SW geometry, we now proceed to  calculate the QNMs frequencies for the extremal RN black hole. To do so, we substitute the dictionary \eqref{eq:negative_branch} into the quantization condition and solve the resulting non-linear algebraic equation for the complex frequency $\omega$. This procedure intrinsically relies on the quantum Matone relation to express the accessory parameter $E$ as a function of the Coulomb branch parameter $a$, the exact expressions for which are detailed in Appendix~\ref{app:ns}.

Since the NS free energy $\mathcal{F}_{\text{inst}}^{(2)}$ is represented as an infinite series expansion in terms of the dimensionless instanton counting parameter $(\Lambda_2 / a)^2$, we must truncate the series at a specific instanton order, denoted by $N_b$. Furthermore, to analytically continue the series beyond its finite radius of convergence and significantly accelerate the numerical evaluation, we apply Pad\'e approximant to the truncated instanton series. 

In all subsequent numerical evaluations, we work in units where the strict extremal RN black hole mass and charge are set to unity ($M=Q=1$), unless otherwise specified.

\subsection{Neutral Massless Scalar Perturbations}
We first focus on neutral massless scalar perturbations  ($q=0, m_p=0$). The numerical solutions for the least-damped QNMs (comprising the fundamental mode $n=0$ and the first few overtones $n=1, 2$) across different angular momentum numbers $l$ are presented in Tables~\ref{tab:l0_qnms_massless}--\ref{tab:l2_qnms_massless}.

As shown in Tables~\ref{tab:l0_qnms_massless}--\ref{tab:l2_qnms_massless}, our quantization condition yields results consistent with the high-precision numerical benchmarks (labeled ``Leaver''). These values, obtained by Onozawa et al.~\cite{Onozawa:1995vu} using a variant of Leaver’s continued fraction method tailored for the extremal horizon, serve as our standard reference. Following the comparative methodology of~\cite{Aminov:2020yma}, we highlight matching significant digits in boldface.

While the continued fraction method is a powerful numerical tool, our geometric approach offers distinct theoretical and analytical advantages. Traditional continued fraction techniques rely on series expansions that naturally fail at the irregular singular point characteristic of the strict extremal limit, necessitating ad-hoc expansions around ordinary points to avoid divergence~\cite{Onozawa:1995vu}. In contrast, the SW/QNM correspondence provides an exact, analytically closed quantization condition derived from the non-perturbative NS free energy. Our framework  absorbs the singular limits of the spacetime geometry into the well-defined decoupling limit of the gauge theory ($N_f=3 \to N_f=2$). Consequently, rather than serving merely as a numerical root-finding algorithm, our approach elucidates the underlying quantum geometric structure of the QNMs spectrum.

For comparison, we also include the traditional the third-order WKB results from the same reference~\cite{Onozawa:1995vu}. A  feature of our gauge-theoretic approach is its systematic improvability. As is visually evident in the Tables, the accuracy of the QNMs frequencies intrinsically depends on the instanton truncation order $N_b$ . By increasing the truncation order from $N_b=2$ to $N_b=12$, the number of matching bold digits systematically increases, demonstrating a rapid convergence of our numerical roots towards the values. This clearly outperforms the traditional WKB approximation, which inherently stagnates at a fixed approximation error (especially for higher overtones $n \ge 1$). 
It is important to emphasize that for higher-order instanton corrections, the application of the Pad\'e approximant is essential; without it, the finite radius of convergence of the instanton series would eventually cause the root-finding routine to diverge.

\footnote{Notably, very recently, Hatsuda and Shiga~\cite{Hatsuda:2026ghx} have employed exact WKB analysis to compute the QNMs of the neutral massless scalar field with unprecedented precision. In their article, they specifically presented a particular situation, $l=0, n=0$ in extremal RN background, and the result is:
\begin{align*}
M\omega =& 0.13345889356706911632231170942642298452384071180138\\
&-0.09584384212811504931055969700757251219004718021629i.
\end{align*}
Our corresponding evaluation (in Table~\ref{tab:l0_qnms_massless}) at the $N_b=12$ instanton order yields $\omega=0.133459 - 0.09584382 i$ (setting $M=Q=1$), which is in agreement with this state-of-the-art exact WKB benchmark.  

It is also important to note that, their exact WKB framework was applied to the neutral massless case. Since for now this analysis is not extended to charged or massive scalar fields yet, a direct comparison with our general results for those scenarios remains unavailable at this level of precision.}

\begin{table}[htbp]
\centering
\renewcommand{\arraystretch}{1.2}
\resizebox{\linewidth}{!}{
\begin{tabular}{@{} l c c c @{}} 
\toprule
$N_b$ & $\omega_0$ ($n=0$) & $\omega_1$ ($n=1$) & $\omega_2$ ($n=2$) \\
\midrule
2   & $\mathbf{0.13}5942 - \mathbf{0.095}31379 i$ & $\mathbf{0.}104219 - \mathbf{0.3}263263 i$ & $\mathbf{0.0}905577 - \mathbf{0.5}722678 i$ \\
6   & $\mathbf{0.133}571 - \mathbf{0.095}79728 i$ & $\mathbf{0.09}30262 - \mathbf{0.330}4667 i$ & $\mathbf{0.0}582712 - \mathbf{0.58}22183 i$ \\
10  & $\mathbf{0.13346}0 - \mathbf{0.09584}264 i$ & $\mathbf{0.0929}726 - \mathbf{0.3306}388 i$ & $\mathbf{0.07}66498 - \mathbf{0.588}9973 i$ \\
12  & $\mathbf{0.1334}59 - \mathbf{0.09584}382 i$ & $\mathbf{0.0929}558 - \mathbf{0.3306}619 i$ & $\mathbf{0.075}4519 - \mathbf{0.588}5324 i$ \\
\midrule 
WKB & $0.121090 - 0.103710 i$ & $0.091570 - 0.337420 i$ & $0.050560 - 0.571640 i$ \\
Leaver & $\mathbf{0.13346} - \mathbf{0.095844} i$ & $\mathbf{0.092965} - \mathbf{0.33065} i$ & $\mathbf{0.075081} - \mathbf{0.58833} i$ \\
\bottomrule
\end{tabular}
}
\caption{Solutions $\omega_n$ to the quantization condition for neutral massless scalar ($q=0, m_p = 0$) perturbations  of the extremal RN black hole with angular momentum $l=0$. The matching digits are highlighted in boldface. The integer $N_b$ denotes the truncation order of the instanton counting series. The WKB and numerical (``Leaver'') values are taken from Ref.~\cite{Onozawa:1995vu} for comparison.}
\label{tab:l0_qnms_massless}
\end{table}

\begin{table}[htbp]
\centering
\renewcommand{\arraystretch}{1.2}
\resizebox{\linewidth}{!}{ 
\begin{tabular}{@{} l c c c @{}} 
\toprule
$N_b$ & $\omega_0$ ($n=0$) & $\omega_1$ ($n=1$) & $\omega_2$ ($n=2$) \\
\midrule
2   & $\mathbf{0.377}546 - \mathbf{0.0}9326217 i$ & $\mathbf{0.34}7954 - \mathbf{0.27}57739 i$ & $\mathbf{0.29}7818 - \mathbf{0.48}55431 i$ \\
6   & $\mathbf{0.377}529 - \mathbf{0.08}381240 i$ & $\mathbf{0.34}7937 - \mathbf{0.276}2973 i$ & $\mathbf{0.298}583 - \mathbf{0.486}0861 i$ \\
10  & $\mathbf{0.3776}10 - \mathbf{0.0893}7425 i$ & $\mathbf{0.34818}4 - \mathbf{0.2761}371 i$ & $\mathbf{0.2984}31 - \mathbf{0.4864}539 i$ \\
12  & $\mathbf{0.3776}33 - \mathbf{0.0893}5923 i$ & $\mathbf{0.34818}1 - \mathbf{0.2761}396 i$ & $\mathbf{0.29846}1 - \mathbf{0.48643}47 i$ \\
\midrule
WKB & $0.375700 - 0.089360 i$ & $0.343920 - 0.278280 i$ & $0.296610 - 0.481450 i$ \\
Leaver & $\mathbf{0.37764} - \mathbf{0.089384} i$ & $\mathbf{0.34818} - \mathbf{0.27614} i$ & $\mathbf{0.29846} - \mathbf{0.48643} i$ \\
\bottomrule
\end{tabular}
}
\caption{Same as Table~\ref{tab:l0_qnms_massless}, but for angular momentum $l=1$.}
\label{tab:l1_qnms_massless}
\end{table}

\begin{table}[htbp]
\centering
\renewcommand{\arraystretch}{1.2}
\resizebox{\linewidth}{!}{
\begin{tabular}{@{} l c c c @{}} 
\toprule
$N_b$ & $\omega_0$ ($n=0$) & $\omega_1$ ($n=1$) & $\omega_2$ ($n=2$) \\
\midrule
2   & $\mathbf{0.62}5186 - \mathbf{0.0}9108756 i$ & $\mathbf{0.60}6390 - \mathbf{0.26}80631 i$ & $\mathbf{0.57}3375 - \mathbf{0.45}77315 i$ \\
6   & $\mathbf{0.62}5179 - \mathbf{0.08}547868 i$ & $\mathbf{0.608}213 - \mathbf{0.269}2497 i$ & $\mathbf{0.5728}39 - \mathbf{0.4582}304 i$ \\
10  & $\mathbf{0.626}367 - \mathbf{0.0}9069463 i$ & $\mathbf{0.608}265 - \mathbf{0.2690}701 i$ & $\mathbf{0.5728}47 - \mathbf{0.45820}82 i$ \\
12  & $\mathbf{0.626}027 - \mathbf{0.088}80593 i$ & $\mathbf{0.608}207 - \mathbf{0.269}1114 i$ & $\mathbf{0.5728}63 - \mathbf{0.458}1931 i$ \\
\midrule
WKB & $0.626090 - 0.088730 i$ & $0.606770 - 0.269440 i$ & $0.572540 - 0.457500 i$ \\
Leaver & $\mathbf{0.62657} - \mathbf{0.088748} i$ & $\mathbf{0.60817} - \mathbf{0.26909} i$ & $\mathbf{0.57287} - \mathbf{0.45820} i$ \\
\bottomrule
\end{tabular}
}
\caption{Same as Table~\ref{tab:l0_qnms_massless}, but for angular momentum $l=2$.}
\label{tab:l2_qnms_massless}
\end{table}

\subsubsection{Pad\'e approximant}
\label{sec:pade}

The exact quantization condition fundamentally relies on the instanton expansion of the NS free energy $\mathcal{F}_{\text{inst}}^{(2)}$. In practice, evaluating the QNMs frequencies involves handling power series expanded in terms of the dimensionless instanton counting parameter. Due to the finite radius of convergence inherent in such non-perturbative corrections, taking these truncated series as standard polynomials can lead to numerical instabilities. This issue becomes pronounced when the instanton truncation order $N_b$ increases, or when exploring higher overtone modes ($n \ge 1$) where the effective expansion parameters are larger.

To analytically continue these series beyond their radius of convergence and ensure the robust evaluation of the quantization condition, we systematically apply the Pad\'e approximant to the relevant series expansions. Specifically, an original truncated series $S(x) = \sum_{k=0}^{N_b} c_k x^k$ is replaced by a rational function $P_M^N(x)$:
\begin{equation}
    P_M^N(x) = \frac{A_0 + A_1 x + \dots + A_M x^M}{1 + B_1 x + \dots + B_N x^N}, \quad M+N \le N_b.
\end{equation}
In our numerical analysis, we consistently utilize diagonal or near-diagonal approximants (i.e., $M \approx N$). Such approximants are well known in the literature for  significantly improving the convergence behavior \cite{Bender:1999box, Baker:1996, Hatsuda:2019eoj, Marino:2015yie}.

\begin{table}[tbp]
\centering
\renewcommand{\arraystretch}{1.3}
\resizebox{\linewidth}{!}{
\begin{tabular}{llcc}
\toprule
Mode ($l=0$) & $N_b$ & With Pad\'e approximant & Without Pad\'e approximant \\
\midrule
\multirow{4}{*}{$n=0$}
& 2-inst & $0.135942 - 0.095314 i$ & $0.135942 - 0.095314 i$ \\
& 10-inst & $\mathbf{0.13346}0 - \mathbf{0.09584}26 i$ & $\mathbf{0.133}664 - \mathbf{0.095}734 i$ \\
& 12-inst & $\mathbf{0.1334}59 - \mathbf{0.09584}38 i$ & $\mathbf{0.133}344 - \mathbf{0.095}921 i$ \\
\cmidrule{2-4}
& Ref.~\cite{Onozawa:1995vu}  & \multicolumn{2}{c}{$0.13346 - 0.095844 i$} \\
\midrule
\multirow{4}{*}{$n=1$}
& 2-inst & $0.104219 - 0.326326 i$ & $0.104219 - 0.326326 i$ \\
& 10-inst & $\mathbf{0.0929}726 - \mathbf{0.3306}39 i$ & $\mathbf{0.09}3947 - \mathbf{0.3}27387 i$ \\
& 12-inst & $\mathbf{0.0929}558 - \mathbf{0.3306}62 i$ & $\mathbf{0.09}3278 - \mathbf{0.33}4537 i$ \\
\cmidrule{2-4}
&Ref.~\cite{Onozawa:1995vu} & \multicolumn{2}{c}{$0.092965 - 0.33065 i$} \\
\midrule
\multirow{4}{*}{$n=2$}
& 2-inst & $0.0905577 - 0.572268 i$ & $0.090558 - 0.572268 i$ \\
& 10-inst & $\mathbf{0.07}66498 - \mathbf{0.588}997 i$ & $0.106999 - 0.635889 i$ \\
& 12-inst & $\mathbf{0.075}4519 - \mathbf{0.588}532 i$ & $0.052947 - 0.578656 i$ \\
\cmidrule{2-4}
& Ref.~\cite{Onozawa:1995vu}  & \multicolumn{2}{c}{$0.075081 - 0.58833 i$} \\
\bottomrule
\end{tabular}
}
\caption{Comparison of the QNMs frequencies evaluated with and without the Pad\'e approximant for the neutral massless ($q=0, m_p = 0$) scalar field at truncation orders $N_b=2,10,12$. The matching digits with the numerical benchmark are highlighted in boldface. Without Pad\'e approximant, higher-order evaluations for overtones (e.g., $n=1, 2$) exhibit significant errors or diverge.}
\label{tab:pade_comparison}
\end{table}

The final physical results are sensitive to this resummation procedure. In Table~\ref{tab:pade_comparison}, we compare the fundamental mode ($n=0$) and the first two overtones ($n=1, 2$) evaluated with and without Pad\'e approximant at various instanton truncation orders. For the fundamental mode, the standard polynomial series express slight deviations at $N_b=12$, whereas the Pad\'e approximant evaluation stabilizes to the benchmark value. Moreover, the necessity of the Pad\'e approximant is much more important to higher overtones. Because the effective expansion parameter becomes larger for excited states, the 10- and 12-instanton results for $n=1$ and $n=2$ without Pad\'e approximant exhibit significant errors or lose physical validity entirely. Conversely, the Pad\'e approximant successfully reconstructs these higher-order corrections, effectively rescuing the series from divergence and maintaining robust convergence toward the exact numerical values. 

\subsection{Neutral Massive Scalar Perturbations }

The introduction of a non-zero scalar mass $m_p$ profoundly alters the effective potential at spatial infinity, lifting it from zero to $m_p^2$.  This asymptotic mass barrier essentially traps low-frequency modes, leading to a phenomenon known as quasi-resonance, where the decay rate of the mode strictly vanishes ($\text{Im}(\omega) \to 0$). Traditional numerical methods face  limitations here: the high-order WKB approximation becomes inaccurate for the fundamental mode ($l=0$) due to the flattening of the potential barrier, while the continued fraction method struggles at the strictly extremal limit ($Q=M$), because the coalescence of horizons merges the regular singularities into an irregular one, rendering standard recurrence relations divergent.

Our non-perturbative quantization condition successfully resolves these mathematical difficulties, providing an analytic handle on the strictly extremal geometry. To demonstrate its robustness, we first anchor our results at higher multipoles where the WKB approximation is relatively reliable. Considering a massive scalar field with $m_p=0.1$ and $l=3$ in the strict extremal limit ($Q=M$), we present the convergence behavior of our quantization condition across different instanton truncation orders ($N_b$) in Table~\ref{tab:l3_convergence}.

\begin{table}[htbp]
    \centering
    \renewcommand{\arraystretch}{1.2}
    \begin{tabular}{lll}
    \toprule
    $N_b\; (Q=M)$ & $\text{Re}(\omega)$ & $-\text{Im}(\omega)$ \\
    \midrule
    2-instanton  & $0.881943$ & $0.095893$ \\
    6-instanton  & $0.875956$ & $0.088647$ \\
    10-instanton & $0.876697$ & $0.089865$ \\
    12-instanton & $0.878275$ & $0.092589$ \\
    \midrule
    WKB ($Q=0.999M$) \cite{Konoplya:2002wt} & $\sim 0.87$ (est.) & $0.088517$ \\
    \bottomrule
    \end{tabular}
    \caption{Calculated QNMs frequencies for a massive neutral ($q=0$) scalar field ($m_p=0.1, l=3, n=0$) in the strict extremal RN limit at different instanton truncation orders $N_b$. The results are compared with the 3rd-order WKB benchmark at the near-extremal limit ($Q=0.999M$) from Ref.~\cite{Konoplya:2002wt}.}
    \label{tab:l3_convergence}
\end{table}

As shown in Table~\ref{tab:l3_convergence}, our evaluation for the strictly extremal background ($Q=M$) yields results that are  consistent with the established WKB estimations. Specifically, our calculated decay rate aligns closely with the traditional third-order WKB benchmark $\text{Im}(\omega) = 0.088517$ reported by Konoplya~\cite{Konoplya:2002wt} for the near-extremal case ($Q=0.999M$). To the best of our knowledge, this specific data point constitutes the only explicitly reported numerical data available in the literature for a neutral massive scalar field in this particular regime; consequently, our comparative analysis is necessarily restricted to this single value, while keeping in mind that the accuracy of  traditional WKB method is limited. Nevertheless, the minor relative difference between the two methods demonstrates a  fundamental consistency. This alignment indicates that our exact geometric approach is practically reliable, and our calculated frequencies---which also furnish the real part ($\text{Re}(\omega) \approx 0.876$) typically absent in standard WKB data---can serve as a useful reference for the QNMs spectrum in the strict extremal limit.

\begin{figure}[htbp]
    \centering
    \begin{subfigure}{0.48\textwidth}
        \includegraphics[width=\textwidth]{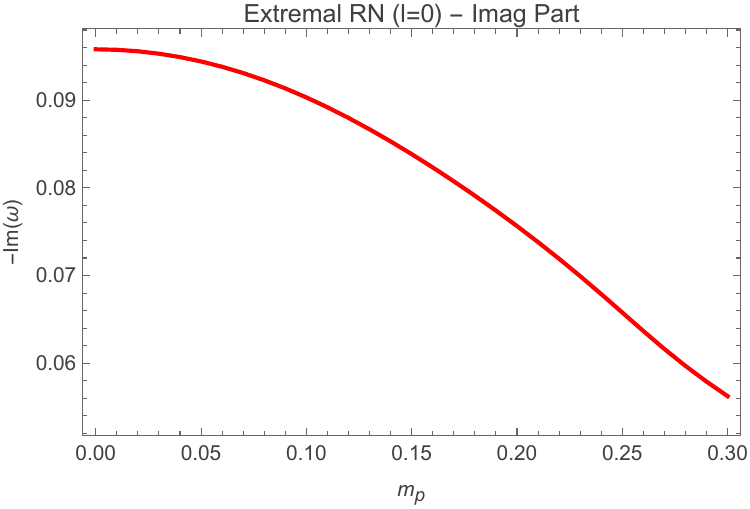}
        \caption{Imaginary part (decay rate)}
        \label{fig:qr_imag}
    \end{subfigure}
    \begin{subfigure}{0.48\textwidth}
        \includegraphics[width=\textwidth]{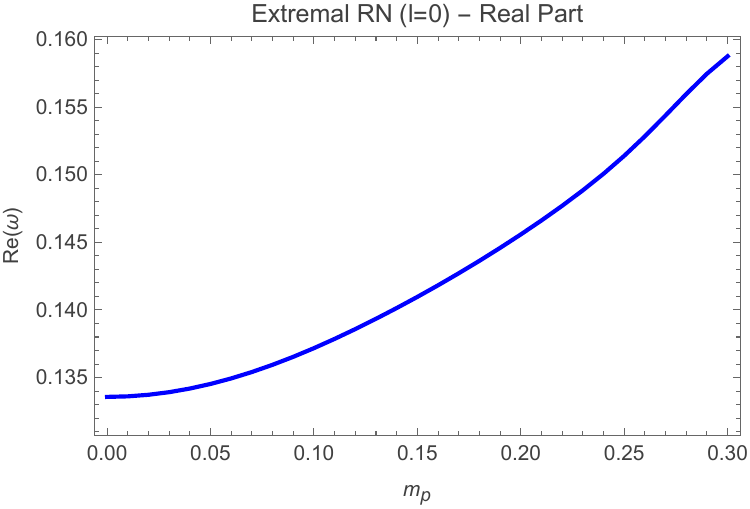}
        \caption{Real part (oscillation frequency)}
        \label{fig:qr_real}
    \end{subfigure}
    \hfill
    \caption{The evolution of the fundamental QNMs ($n=0, l=0$) for a neutral ($q=0$) scalar field as a function of the dimensionless mass $m_p$ in the strict extremal RN limit. The decay rate strictly approaches zero, explicitly illustrating the transition into the quasi-resonance regime.}
    \label{fig:quasiresonance}
\end{figure}

Furthermore, our analytic framework allows us to smoothly track the evolution of the fundamental mode ($l=0$) across a wide range of scalar masses. As illustrated in Figure~\ref{fig:quasiresonance}, we present the continuous trajectory of both the imaginary and real parts of the $l=0$ mode as $m_p$ increases. A notable physical feature is observed in Figure~\ref{fig:qr_imag}: as the dimensionless mass parameter $m_p$ increases, the decay rate $-\text{Im}(\omega)$ is non-linearly suppressed and monotonically approaches zero. This continuous trajectory of the decay rate precisely replicates the physical trend observed by Ohashi and Sakagami \cite{Ohashi:2004wr} for near-extremal black holes ($Q=0.99M$) using generalized continued fractions. However, our result is obtained analytically at the exact $Q=M$ limit. Furthermore, our explicit visualization of the real part evolution (Figure~\ref{fig:qr_real}) directly complements the parametric complex-plane trajectories presented in earlier numerical literature. Physically, this confirms that for sufficiently large $m_p$, the scalar wave is completely confined between the infinite $AdS_2$ throat near the degenerate horizon and the $m_p^2$ barrier at spatial infinity. The SW geometry thus captures the cutoff of black hole dissipation, providing a description of quasi-resonances without suffering from the singularity degeneracies that plague standard numerical schemes.

\subsection{Charged Massless Scalar Perturbations}
\label{sec:charged_massless}

The extension of our framework to charged scalar fields ($q \neq 0$) offers an alternative analytical perspective when evaluating the exactly extremal spectrum. As noted in the literature~\cite{Richartz:2014jla}, standard numerical root-finding algorithms face challenges due to the coalescence of the inner and outer horizons ($r_+ \to r_-$). Mathematically, Leaver's traditional continued fraction method relies on a Frobenius expansion tailored to the regular singular structure of the CHE. In the strict extremal limit, the collision of these two regular singularities forms an irregular singularity of Poincar\'e rank 1, characterizing the DCHE. This topological singularity coalescence causes the convergence radius of the standard Frobenius series to strictly vanish, rendering the corresponding recurrence relations divergent.

To address this irregular singularity issue, specialized numerical techniques have been developed. Notably, Onozawa et al.~\cite{Onozawa:1995vu} modified the continued fraction method for neutral perturbations. This technique was later generalized by Richartz \cite{Richartz:2015saa} to incorporate charged massless fields. 

In this context, our $N_f = 2$ SW geometric approach provides a natural resolution. By intrinsically mapping the exact radial equation directly to the DCHE, this framework naturally accommodates the singularity coalescence. It bypasses the divergent local expansion and constructs a global solution framework. 

To demonstrate its accuracy, we compare our 12-instanton evaluation for charged massless fields ($m_p=0$) with the numerical benchmarks obtained by Richartz \cite{Richartz:2015saa}. As shown in Table \ref{tab:charged_massless}, our exact non-perturbative quantization accurately reproduces the high-precision numerical roots for various $\ell$ and field charges, verifying the robustness of the SW/QNM dictionary in the charged regime.

\begin{table}[htbp]
\centering
\renewcommand{\arraystretch}{1.3}
\begin{tabular}{lccc}
\toprule
$\ell$ & $q$ & Our Results (12-instanton) & Numerical Benchmark \cite{Richartz:2015saa} \\
\midrule
\multirow{2}{*}{$\ell=0$} & $0.01$ & $0.138478 - 0.095816i$ & $0.138478 - 0.095816i$ \\
                          & $0.1$  & $0.185411 - 0.093018i$ & $0.185411 - 0.093018i$ \\
\midrule
\multirow{2}{*}{$\ell=1$} & $0.01$ & $0.382648 - 0.089354i$ & $0.382657 - 0.089379i$ \\
                          & $0.1$  & $0.429127 - 0.088813i$ & $0.429135 - 0.088838i$ \\
\midrule
\multirow{2}{*}{$\ell=2$} & $0.01$ & $0.631037 - 0.088804i$ & $0.631582 - 0.088746i$ \\
                          & $0.1$  & $0.676994 - 0.088594i$ & $0.677534 - 0.088542i$ \\
\bottomrule
\end{tabular}
\caption{Comparison of the fundamental QNMs frequencies ($n=0$) for charged massless ($m_p = 0$) scalar fields at strict extremality. Our 12-instanton evaluations are in agreement with the numerical benchmarks.}
\label{tab:charged_massless}
\end{table}

\subsection{Charged Massive Scalar Perturbations}
\label{sec:charged_massive}

While Leaver's method successfully resolved the charged massless case, its application to massive probes ($m_p \neq 0$), particularly those that are simultaneously charged and massive, remains elusive. Previous studies often relied on near-extremal approximations, which is difficult to capture the precise full spectrum at strict extremality. A unified, global non-perturbative evaluation for simultaneously charged and massive fields therefore remains an open challenge.

By intrinsically mapping the exact radial equation to the DCHE, our $N_f=2$ SW geometric framework is naturally immune to this singularity confluence, allowing us to extract the true discrete global QNMs. To demonstrate the convergence of our approach, we first investigate the most intricate regime involving both electrostatic and mass barriers ($m_p \neq 0, q \neq 0$). 

In Table \ref{tab:charged_convergence}, we present the exact convergence behavior of our quantization condition for a massive charged scalar field ($m_p=0.1, \ell=3, n=0$) across different instanton truncation orders $N_b$, evaluated with Pad\'e approximant, and compared with the  third order WKB results of the nearly extremal RN black holes ($Q=0.999M$) in the literature~\cite{Konoplya:2002wt}.

\begin{table}[htbp]
\centering
\renewcommand{\arraystretch}{1.3}
\begin{tabular}{lcc}
\toprule
$N_b$ ($m_p=0.1, \ell=3, n=0$) & $\omega$ ($q=0.05$) & $\omega$ ($q=0.1$) \\
\midrule
2-instanton  & $0.907079 - 0.095879i$ & $0.932555 - 0.095813i$ \\
6-instanton  & $0.901093 - 0.088633i$ & $0.926580 - 0.088567i$ \\
10-instanton  & $0.901834 - 0.089852i$ & $0.927319 - 0.089784i$ \\
12-instanton & $0.903412 - 0.092574i$ & $0.928892 - 0.092498i$ \\
\midrule
3rd-order WKB ($Q=0.999M$) & $\sim \text{est.} - 0.088512i$ & $\sim \text{est.} - 0.088470i$ \\
\bottomrule
\end{tabular}
\caption{Convergence of the fundamental QNMs frequencies for a charged massive scalar field at strict extremality. The 6-instanton evaluation closely matches the near-extremal WKB benchmark (in ref.\cite{Konoplya:2002wt}), while the 12-instanton evaluation provides the exact converged values and the previously absent real frequencies.}
\label{tab:charged_convergence}
\end{table}

Having established the convergence of our framework, we broaden our analysis to demonstrate its universal applicability. In Table~\ref{tab:charged_grid}, we present an extended parameter scan of the QNMs spectrum evaluated at the 12-instanton order, covering various representative values of scalar masses ($m_p$), charges ($q$), multipoles ($\ell$), and overtones ($n$).  
The dashes in Table~\ref{tab:charged_grid} denote regions of numerical instability arising from specific parameter combinations; a detailed investigation of this regime is deferred to future work.

\begin{table}[htbp]
\centering
\renewcommand{\arraystretch}{1.3}
\resizebox{\linewidth}{!}{
\begin{tabular}{lcccc}
\toprule
$/$ & $q=0.05$ & $q=0.1$ & $q=0.15$ & $q=0.2$ \\
\midrule
$m_p=0$ & $0.158947-0.095142i$ & $0.185411-0.093018i$ & $0.212849-0.089415i$ & $0.241254 - 0.084226i$ \\
\midrule
$m_p=0.1$ & $0.163288 - 0.090459i$ & $0.190241 - 0.089268i$ & $0.217977 - 0.086614i$ & $0.24654 - 0.082366i$ \\
$m_p=0.2$ & $0.173868 - 0.077009i$ & $0.202663 - 0.077713i$ & $0.231799 - 0.077483i$ & --\\
$m_p=0.3$ & $0.187928 - 0.058159i$ & $0.218389 - 0.060460i$ & $0.249428 - 0.062150i$ & -- \\
\bottomrule
\end{tabular}
}
\caption{Parameter scan of the strictly extremal QNMs frequencies for charged scalar fields at $\ell=0, n=0$, evaluated at the 12-instanton order with Pad\'e approximant. 
}
\label{tab:charged_grid}
\end{table}

In the absence of existing high-precision benchmarks for the fully charged and massive case, verifying the numerical stability of our evaluations is of paramount importance. Similar to the systematic convergence behavior demonstrated in Section~\ref{sec:numerical} (e.g., Tables~\ref{tab:l0_qnms_massless}-\ref{tab:l2_qnms_massless}), we have rigorously tested the instanton truncation dependence for these novel predictions. The quasinormal frequencies reported in Table~\ref{tab:charged_grid}, evaluated at the $N_b=12$ order with Pad\'e approximant, exhibit robust stabilization. Specifically, the frequency shifts between the $N_b=10$ and $N_b=12$ truncations are confined to the last few reported significant digit (mostly the last digit),  see in appendix~\ref{app:large}. This internal consistency confirms the reliability of our analytical framework in extracting the global discrete QNMs spectrum within this previously inaccessible regime.  

\section{Conclusion and Outlook}
\label{sec:conclusion}

In this paper, we have extended the SW/QNM correspondence to the extremal RN black hole in asymptotically flat spacetime. Our main analytical result is the exact construction of the dictionary between the scalar perturbations of the extremal RN black hole and the parameters of the quantum $N_f=2$ SW curve. Based on this mapping, we derived the geometric reductions of the radial equation at two critical physical thresholds. Specifically, we demonstrated that the superradiance limit ($\omega \to q$) and the mass decoupling limit ($\omega \to m_p$) force the DCHE to degenerate into the Whittaker equation and the RDCHE, respectively. Remarkably, our exact dictionary flawlessly translates these geometric reductions into the canonical weak-coupling limit and the flavor-decoupling flow in the corresponding supersymmetric gauge theory.

Numerically, we verified the robustness of our quantization condition. For both neutral and charged massless scalar perturbations, the evaluation of the NS free energy yielded resonant frequencies in agreement with state-of-the-art numerical benchmarks. More significantly, for neutral massive scalars ($m_p \neq 0$), our analytic framework successfully tracked the fundamental mode ($\ell=0$) into the quasi-resonance regime, explicitly illustrating the cutoff of dissipation ($\text{Im}(\omega) \to 0$) due to the asymptotic mass barrier. This demonstrates the method's capability to resolve the singularity coalescence of the DCHE arising at the strict extremal limit, where standard numerical techniques may face limitations. 

Building on this, we provided the first 
predictions for the discrete global QNMs spectrum of simultaneously charged and massive scalar fields at strict extremality. Furthermore, our 12-instanton evaluation successfully extracted the precise real oscillation frequencies that were previously absent in asymptotic analyses.

Looking ahead, finding a systematic analytic approach to resolve the QNMs spectrum problems in more situations remains an intriguing open question. Another promising direction is to make this framework mathematically accommodating the distinct asymptotic behavior of quasi-bound states. We plan to systematically explore these physically rich boundaries and spectra in future investigations.

\acknowledgments
The authors thank Prof. Xian-Hui Ge, Rui-Dong Zhu,  Hong-Fei Shu, Yang Lei and Masataka Matsumoto for helpful discussions.  K.Z. (Hong Zhang) thanks Prof. Hong L\"u for drawing attention to this interesting subject.
We are particularly grateful to the anonymous referee for their insightful comments and highly constructive suggestions, which directly prompted the extension of our analysis to the charged massive case  and significantly broadened the theoretical scope of this manuscript. This work is supported by a classified fund from Shanghai city.

\appendix
\section{NS Free Energy for $N_f=2$}
\label{app:ns}

For self-containment, this appendix briefly reviews the exact definitions and the instanton expansion of the NS free energy for the four-dimensional $\mathcal{N}=2$ $\mathrm{SU(2)}$ gauge theory coupled to $N_f=2$ fundamental hypermultiplets. The combinatorial formulations and notation adopted here follow the standard literature, particularly relying on the conventions established in~\cite{Aminov:2020yma}.

\subsection{Combinatorial Formulation}

The NS limit of the free energy is extracted from the full instanton partition function $\mathcal{Z}$ by taking the specific $\Omega$-background limit $\epsilon_2 \to 0$ while keeping $\epsilon_1 \equiv \hbar$ finite:
\begin{equation}
    F_{\text{inst}}^{(N_f)}(a; \mathbf{m}; \Lambda_{N_f}, \hbar) = \lim_{\epsilon_2 \to 0} \epsilon_2 \log \mathcal{Z}^{(N_f)}(ia, \mathbf{m}, \hbar, \epsilon_2).
\end{equation}
Here, the parameters $a$, $\mathbf{m} = \{m_1, m_2\}$, and $\Lambda_2$ denote the Coulomb branch vacuum expectation value (acting as the corresponding K\"ahler modulus), the hypermultiplet masses, and the dynamical scale of the theory, respectively. 

Based on the standard localization techniques, the general $\Omega$-deformed $\mathrm{U(2)}$ Nekrasov partition function for $N_f=2$ is evaluated as a sum over pairs of Young partitions $\mathbf{Y} = (Y_1, Y_2)$:
\begin{equation}
    \mathcal{Z}^{(2)}\left(a;\mathbf{m};\Lambda_{2},\epsilon_{1},\epsilon_{2}\right)=\sum_{\mathbf{Y}}\left(\frac{\Lambda_{2}^{2}}{4}\right)^{\ell(\mathbf{Y})}\mathcal{Z}_{\mathbf{Y}}^{\text{gauge}}\mathcal{Z}_{\mathbf{Y}}^{\text{matter}},
\end{equation}
where $\ell(\mathbf{Y}) = |Y_1| + |Y_2|$ counts the total number of boxes in the partition pair. Setting the Coulomb branch parameters to $\alpha_{2}=-\alpha_{1}=a$, the vector multiplet (gauge) contribution reads:
\begin{equation}
\begin{aligned}
\mathcal{Z}_{\mathbf{Y}}^{\text{gauge}} 
&= \prod_{I,J=1}^{2}\; 
\prod_{s\in Y_{I}} 
\frac{1}{\alpha_{I}-\alpha_{J}
      -\epsilon_{1}\,v_{Y_{J}}(s)
      +\epsilon_{2}\bigl(h_{Y_{I}}(s)+1\bigr)} \\
&\qquad\times 
\prod_{s\in Y_{J}} 
\frac{1}{\alpha_{I}-\alpha_{J}
      +\epsilon_{1}\bigl(v_{Y_{I}}(s)+1\bigr)
      -\epsilon_{2}\,h_{Y_{J}}(s)},
\end{aligned}
\end{equation}
while the contribution originating from the $N_f=2$ fundamental matter multiplets is:
\begin{equation}
\mathcal{Z}_{\mathbf{Y}}^{\text{matter}}
= \prod_{k=1}^{2}
  \prod_{I=1}^{2}
  \prod_{(i,j)\in Y_{I}}
  \Bigl(\alpha_{I}+m_{k}
        +\bigl(i-\tfrac12\bigr)\epsilon_{1}
        +\bigl(j-\tfrac12\bigr)\epsilon_{2}\Bigr).
\end{equation}
For any specific box $s=(i,j)$ located within a Young diagram $Y$, its arm length $h_Y(s)$ and leg length $v_Y(s)$ are given by $h_{Y}(s) = y_{i} - j$ and $v_{Y}(s) = y_{j}^t - i$, where $y_i$ and $y_j^t$ specify the lengths of the $i$-th row and the $j$-th column, respectively.

\subsection{Full Free Energy and Instanton Expansion}

The quantization condition fundamentally relies on the derivative of the complete NS free energy with respect to $a$. This full expression systematically incorporates the classical, one-loop, and non-perturbative instanton terms:
\begin{equation}
\begin{split}
    \partial_{a} \mathcal{F}^{(2)}(a ; \mathbf{m} ; \Lambda_2, \hbar) =\, & -4 a \log \left(\frac{\Lambda_2}{2 \hbar}\right) - \pi \hbar - 2 \mathrm{i} \hbar \log \left[ \frac{\Gamma\left(1+\frac{2 \mathrm{i} a}{\hbar}\right)}{\Gamma\left(1-\frac{2 \mathrm{i} a}{\hbar}\right)} \right] \\
    & - \mathrm{i} \hbar \sum_{j=1}^{2} \log \left[ \frac{\Gamma\left(\frac{1}{2}+\frac{m_{j}-\mathrm{i} a}{\hbar}\right)}{\Gamma\left(\frac{1}{2}+\frac{m_{j}+\mathrm{i} a}{\hbar}\right)} \right] + \frac{\partial \mathcal{F}_{\text{inst}}^{(2)}(a ; \mathbf{m} ; \Lambda_2, \hbar)}{\partial a} \,.
\end{split}
\end{equation}

To correctly describe the $\mathrm{SU(2)}$ gauge theory, the spurious $\mathrm{U(1)}$ factor must be decoupled from the $\mathrm{U(2)}$ partition function \cite{Aminov:2020yma}. This decoupling modifies the first instanton correction, yielding the true $\mathrm{SU(2)}$ instanton free energy:
\begin{equation}
    \mathcal{F}_{\text{inst}}^{(2)}(a;\mathbf{m};\Lambda_{2},\hbar)=F_{\text{inst}}^{(2)}(a;\mathbf{m};\Lambda_{2},\hbar)-\frac{\Lambda_{2}^{2}}{8}.
\end{equation}
Evaluating the combinatorial sums, the power series expansion of $F_{\text{inst}}^{(2)}$ in the dynamical scale $\Lambda_2$, evaluated up to $\mathcal{O}(\Lambda_2^6)$, explicitly reads:
\begin{equation}
\label{eq:F_inst_expansion}
\begin{split}
    F_{\text{inst}}^{(2)}(a;\mathbf{m};\Lambda_{2},\hbar) =\, & \left( \frac{1}{2} - \frac{(4m_1 m_2)}{2(4a^2+\hbar^2)} \right) \frac{\Lambda_2^2}{4} \\
    & - \frac{\Lambda_2^4}{1024(a^2+\hbar^2)(4a^2+\hbar^2)^3} \bigg( 64a^2 \big( a^4+3a^2(m_1^2+m_2^2)+5m_1^2m_2^2 \big) \\
    &+ \hbar^6 + 12\hbar^4(a^2+m_1^2+m_2^2) + 16\hbar^2 \big( 3a^4+6a^2(m_1^2+m_2^2)\\
    &-7m_1^2m_2^2 \big) \bigg)  + \mathcal{O}(\Lambda_2^6) \,.
\end{split}
\end{equation}

\subsection{Quantum Mirror Map via Quantum Matone Relation}

The gauge theoretic parameter $a$ is implicitly related to the energy eigenvalue $E$ through the exact quantum mirror map. For $\mathrm{SU(2)}$ theories, this is efficiently encoded by the quantum version of the Matone relation \cite{Flume:2004rp, Aminov:2020yma}. Specializing to the $N_f=2$ case, the exact relation states:
\begin{equation}
    E = a^2 - \frac{1}{2} \Lambda_2 \partial_{\Lambda_2} \mathcal{F}_{\text{inst}}^{(2)}.
\end{equation}
During the numerical procedure, we invert this relation perturbatively using the explicit expansion \eqref{eq:F_inst_expansion}. This provides the necessary analytic continuation to express $a$ strictly as a function of the physical spectral variables $(E, m_1, m_2, \Lambda_2)$, which is subsequently implemented into the Bohr-Sommerfeld quantization condition.

\newpage
\section{Datas for Selected Charged Massive Scalar Perturbations}
\label{app:large}
\begin{table}[htbp]
\centering
\resizebox{\textwidth}{!}{
\renewcommand{\arraystretch}{1.2}
\begin{tabular}{c c c c c c}
\toprule
$m_p$ & $q$ & $N_b=2$ (2-inst) & $N_b=6$ (6-inst) & $N_b=10$ (10-inst) & $N_b=12$ (12-inst) \\
\midrule
$0$ & $0.05$ & $0.161401 - 0.0946931i$ & $0.159058 - 0.0951024i$ & $0.158948 - 0.0951408i$ & $0.158947 - 0.0951418i$ \\
$0$ & $0.1$ & $0.187774 - 0.092813i$ & $0.185519 - 0.0929991i$ & $0.185412 - 0.0930173i$ & $0.185411 - 0.0930178i$ \\
$0$ & $0.15$ & $0.215036 - 0.0896179i$ & $0.212946 - 0.0894285i$ & $0.212851 - 0.0894148i$ & $0.212849 - 0.0894146i$ \\
$0$ & $0.2$ & $0.243107 - 0.085005i$ & $0.241321 - 0.0842788i$ & $0.241255 - 0.0842267i$ & $0.241254 - 0.0842258i$ \\
\midrule
$0.1$ & $0.05$ & $0.165467 - 0.0903961i$ & $0.163383 - 0.0904519i$ & $0.16329 - 0.0904587i$ & $0.163288 - 0.090459i$ \\
$0.1$ & $0.1$ & $0.192309 - 0.0891984i$ & $0.190328 - 0.0892606i$ & $0.190242 - 0.0892679i$ & $0.190241 - 0.0892682i$ \\
$0.1$ & $0.15$ & $0.219899 - 0.086716i$ & $0.218055 - 0.0866187i$ & $0.217978 - 0.086614i$ & $0.217977 - 0.0866141i$ \\
$0.1$ & $0.2$ & $0.248248 - 0.0828294i$ & $0.246604 - 0.0823938i$ & $0.246541 - 0.0823662i$ & $0.24654 - 0.0823658i$ \\
\midrule
$0.2$ & $0.05$ & $0.175134 - 0.0783312i$ & $0.173883 - 0.0770785i$ & $0.173868 - 0.0770098i$ & $0.173868 - 0.0770087i$ \\
$0.2$ & $0.1$ & $0.204104 - 0.0783593i$ & $0.202709 - 0.0777486i$ & $0.202664 - 0.0777134i$ & $0.202663 - 0.0777128i$ \\
$0.2$ & $0.15$ & $0.233212 - 0.0777481i$ & $0.231847 - 0.0774953i$ & $0.2318 - 0.0774828i$ & $0.231799 - 0.0774827i$ \\
\midrule
$0.3$ & $0.05$ & $0.177936 - 0.068034i$ & $0.18832 - 0.0579548i$ & $0.187931 - 0.0581713i$ & $0.187928 - 0.0581588i$ \\
$0.3$ & $0.1$ & $0.216169 - 0.0636182i$ & $0.2183 - 0.0603313i$ & $0.218391 - 0.0604574i$ & $0.218389 - 0.0604598i$ \\
$0.3$ & $0.15$ & $0.249611 - 0.0638605i$ & $0.24938 - 0.0621882i$ & $0.249427 - 0.0621497i$ & $0.249428 - 0.0621496i$ \\
\bottomrule
\end{tabular}
} 
\vspace{0.2cm}
\caption{Full truncation dependence of the fundamental QNMs frequencies for various combinations of scalar mass $m_p$ and charge $q$. This table details the results obtained under the Pad\'e approximant scheme at truncation orders $N_b$=2, 6, 10, and 12 for all entries not marked by dashes in Table~\ref{tab:charged_grid}, underscoring the robust convergence across these orders.}
\label{tab:full_breakdown}
\end{table}

\setcounter{footnote}{0}

\bibliographystyle{JHEP}
\bibliography{RN_QNM}

\end{document}